\documentclass[namedreferences]{solarphysics}

\usepackage[hyperref,optionalrh]{spr-sola-addons} 
\usepackage{graphicx}        
\usepackage{color}           

\chardef\us=`\_

\begin{document}

\begin{article}
\begin{opening}

\title{Correlation Functions of Photospheric Magnetic Fields in Solar Active Regions   }

\author[addressref=aff1,corref,email={vabramenko@gmail.com}]{\inits{V.}\fnm{Valentina}~\lnm{Abramenko}}
\author[addressref={aff1},email={bictr97@gmail.com}]{\inits{R.}\fnm{Regina}~\lnm{Suleymanova}}

\address[id=aff1]{Crimean Astrophysical Observatory of Russian Academy of Sciences, Nauchny 298409,
	Bakhchisaray, Republic of Crimea}

\runningauthor{V. Abramenko and R. Suleymanova }
\runningtitle{\textit Correlation Functions of Solar Active Regions}

\begin{abstract}
We used magnetograms acquired with the {\it Helioseismic and Magnetic Imager} (HMI) on board the {\it Solar Dynamics Observatory} (SDO) to calculate and analyze spatial correlation functions and the multi-fractal spectra in solar active regions (ARs). The analysis was performed for two very different types of ARs: i) simple bipolar magnetic structures with regular orientation (the magneto-morphological class A1), and ii) very complex multi-polar ARs (the magneto-morphological class B3). All ARs were explored at the developed phase during flareless periods. For correlation functions, the power-law and exponential approximations were calculated and compared. It was found that the exponential law holds for the correlation functions of both types of ARs within spatial scales of 1-36~Mm, while the power law failed to approximate the observed correlation functions. The property of multi-fractality was found in all ARs, being better pronounced for the complex B3-class ARs. Our results might imply that photospheric magnetic fields of an AR is a self-organized system, which, however, does not exhibit properties of self-organized criticality (SOC), and its fractal properties are an attribute of more broad (than SOC only) class of non-linear systems. 

\end{abstract}
\end{opening}

\section{Introduction}
\label{S-Introduction} 
     An active region (AR) on the Sun is a large cluster of magnetized plasma embracing a volume from sub-photospheric depths through the photosphere to the chromosphere and the corona. The magnetic field is a structuring agent that forms a visible appearance of an AR. The AR evolves amid a turbulent medium, and therefore, as any turbulent phenomenon, evolves as a non-linear dynamical dissipative system, possessing properties of self-organization,  see, e.g.,  \citet{Biskamp1993, Frisch1995, Kurakin2011, Aschwanden2018}). 
         
      \citet{Bak1987} introduced a new concept of self-organized criticality (SOC). A self-organized system, when it is in a SOC state, can spontaneously transit into a critical state in which a catastrophe of any scale (up to the scale of the entire system) might occur. To this end, SOC-systems form a subset of self-organized systems. According to \citet{Watkins2016}, one way to detect the SOC-systems is to analyze their correlation functions. In the SOC state a system is capable of inducing long-range correlations meaning that fluctuations at all scales up to the longest ones are possible, resulting in a heavy-tailed power law correlation functions. On the contrary, a self-organized systems that exhibit exponential correlation functions are not in the SOC state. 
      
      Therefore, an analysis of correlation functions may help us to reveal properties of self-organization versus SOC properties. We note that the correlation functions of 2D-structures of solar magnetic fields were not extensively studied.  Except for our old publication (\cite{Abramenko2003}), we found only one publication for the last two decades, where the 2D correlation function technique was applied to analyze the spatial correlation of magnetic field fluctuations:  \cite{Baumgartner2022}.
      Here we attempt to fill the gap and to explore the correlation functions of the radial magnetic field component in ARs using space-born data. 
      The main aim of the present pilot study is to derive the spatial correlation function of the magnetic field in an AR, leaving for future the important question on the possible changes in this function related to the evolution of the AR. For our purpose, the best way would be to explore the AR at the time without lateral circumstances such as emergence, decay, or flare, i.e., during the flareless interval of the developed phase.    
      
      This study is a case study, and the outcome might be better understood if the analysis involves contrast cases. We thus selected four ARs of extremely simple and regular magnetic configuration and compared their properties with those of four ARs of very complex and irregular configuration (Sec. \ref{S-CF}). An accompanying analysis of fractal properties of all ARs are presented in Sec. \ref{S-Fr}, and our concluding remarks are gathered in Sec. \ref{S-Con}.    
     
 \section{Data}
 \label{S-Data} 
 
We used magnetograms acquired with the {\it Helioseismic and Magnetic Imager } (HMI) on board the {\it Solar Dynamics Observatory} (SDO) \cite{Scherrer2012}, \cite{Schou2012}. To retrieve the data, we downloaded {\it Space-weather HMI Active Region Patches} (SHARP, sharp\_cea\_720s series) from the {\it Joint Science Operations Center} (JSOC, \url{http://jsoc.stanford.edu/}). The $B_r$, $B_p$, and $B_t$ magnetograms were acquired in the Fe I 6173.3\AA\ spectral line with the spatial resolution of 1~arcsec. The radial component, $B_r$, was utilized in the present research. 
   
\begin{figure}
\centerline{\includegraphics[width=1.0\textwidth,clip=]{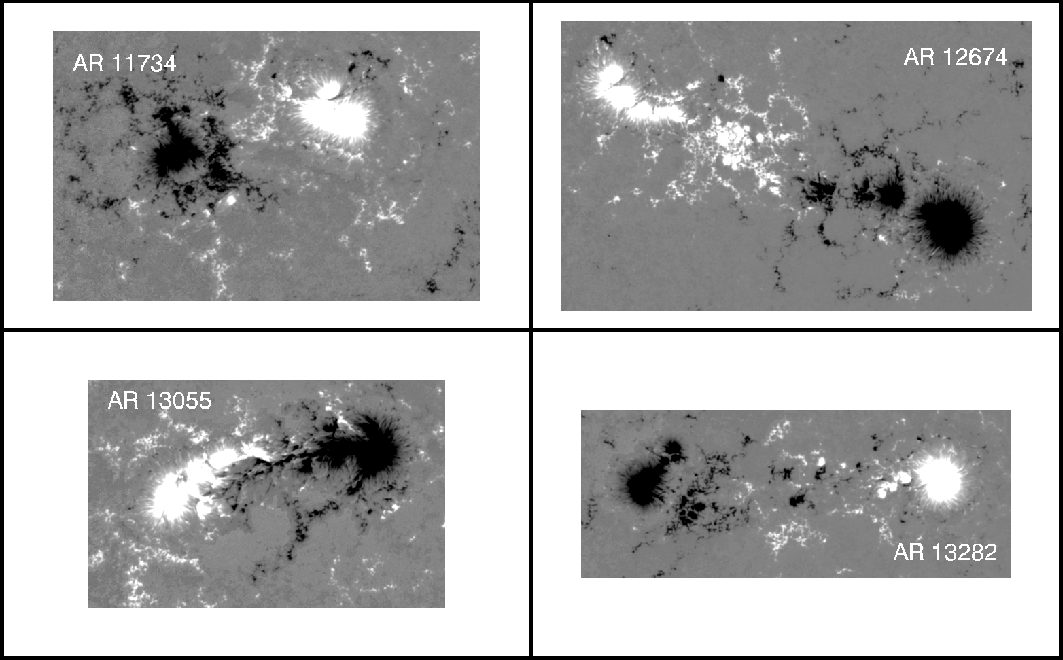}}
\caption{ SDO/HMI magnetograms of four regular ARs (MMC-class A1). All magnetograms are of the same spatial scale, and the horizontal dimension of the AR 11734 panel is 228 Mm (629 pixels). The magnetograms are scaled from -800 Mx$\cdot$cm$^{-2}$ (black) to 800 Mx$\cdot$cm$^{-2}$ (white). East is to the left, North is to the top. The solar equator is parallel to the horizontal border of each magnetogram.}\label{fig1}
\end{figure}

\begin{figure}
	\centerline{\includegraphics[width=1.0\textwidth,clip=]{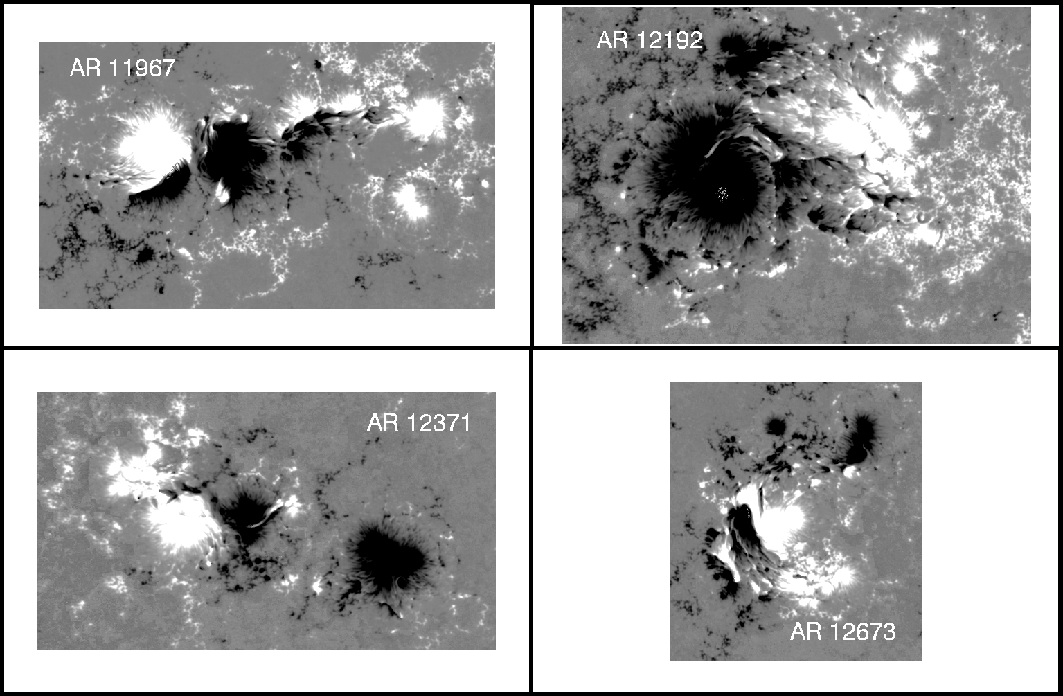}
	}
\caption{SDO/HMI magnetograms of four irregular ARs (MMC-class B3).  All magnetograms are of the same spatial scale, the E-W length of the AR 11967 is 239 Mm (659 pixels). Other notations are the same as in Figure \ref{fig1}.  }
	\label{fig2}
\end{figure}

In total, eight ARs were chosen for the study (see Table \ref{Table1} and Figures \ref{fig1}, \ref{fig2}). 
Our criteria for selection can be summarized as follows.
	Generally, statistical properties of the photospheric magnetic field (e.g.,  correlation functions, structure functions,  distribution functions, statistical moments, etc. ) can vary during the evolution of the AR. The question deserves a separate investigation, and in the present study, to begin with, we restricted ourselves to the developed phase of the AR’ evolution. So, all ARs in Table \ref{Table1} are mature ARs, observed in the developed phase. 
	A choice of the AR’s position on the solar disc was governed by two considerations. First, we tried to select the observation day when the AR is close to the central meridian to mitigate the errors for the projection. Second, we tried to explore the AR during the first half of the developed phase, when the destroy process did not start to corrupt the magnetic structure. As we mentioned above, the decay process analysis was postponed for future.  We managed to comply with these restrictions for all ARs except for the last one,  AR NOAA 12673, which started a fast development being in the Western hemisphere.  
	A possible influence of flaring on the statistical functions of the magnetic field is worthy of a special analysis. So, in the present pilot study, we tried to avoid flaring, and so, all taken magnetograms were recorded during non-flaring periods and the strongest flare in an AR occurred a day(s) before (negative delay in the last column of Table \ref{Table1}), or after (positive delay)  the selected interval of analysis. 

The selection was also guided by the magneto-morphological classification (MMC) of ARs, suggested in \citet{Abramenko2018,Abramenko2021}. Following the MMC, four chosen ARs are of the A1-class, which includes simple bipolar ARs (see Figure \ref{fig1}) complying with the empirical laws (the Hale polarity law, the Joy's law, the prevalence of the leading spot).  We refer to them as regular ARs. The remainig four ARs (Figure \ref{fig2}) belong to the B3-class comprised of the most complex ARs caused, presumably, by emergence of several inter-twinned flux tubes. To ease the comparison, all selected ARs had carried a large amount of the total magnetic flux (in excess of 2.7$\cdot 10^{22}$ Mx, see 4th column in Table \ref{Table1}). 
Note that the total unsigned magnetic flux of an AR was calculated as a sum of absolute values of the magnetic flux density $B_r$ in pixels where $|B_r| > $18 Mx cm$^{-2}$, multiplied by the pixel size. The threshold magnitude was derived as a standard deviation of $B_r$ in a quiet-sun area. As soon as the flux values are used here for the illustration only, the choice of the threshold does not affect the results.
     
The choice of A1- and B3-class ARs allowed us to take into account possible differences in the sub-photospheric origin of the ARs: A1 class ARs are thought to be the most compliant with the mean field dynamo theory, where active region are thought to be formed from the coherent toroidal magnetic field. The B3-class ARs appear to be strongly affected by the sub-photospheric turbulent convection and thus most strongly deviate from the mean field dynamo theory predictions \citep{Abramenko2021}. It was shown earlier that the magnetic flux from A- and B-class ARs varies differently over the solar cycle \citep{Abramenko2023}: while the A-class ARs determine the overall shape of the cycle profile, the B-class ARs determine the multi-peak structure of the cycle maximum. During the solar minima predominantly A-class ARs appear. The A- and B-class ARs show different flaring activity as well: powerful X-class flares and GLE-events registered by neuron monitors are predominantly associated with B2 and B3-class ARs \citep{Abramenko2021,Suleymanova2024}.

Flaring productivity of an AR was quantified here with the flare index (FI,
\cite{Abramenko2005b}), which was derived by summing the GOES-class
of all flares observed in an AR during its passage across the solar
disk, $\tau$, and then normalizing the total by $\tau$. Further scaling was
applied so that an AR with one C1.0 (X1.0) flare per day has the
flare index FI=1.0 (100).

The dissimilarity between the A1-class and B3-class ARs in flaring activity is demonstrated in the last column of Table \ref{Table1}, where the flare index, FI, the strongest flare launched by an AR during its passage across the solar disk, and the time delay between the strongest flare and the observation interval boundary are listed against the AR class. The flare index indicates that the flare productivity of the B3-class ARs is approximately by two orders of magnitude higher than that of the A1-class ARs.

Thus, if any difference exists in the correlation functions, it should be revealed when comparing the A1- and B3-class ARs.       
     
    \begin{table}
    	\caption{Active regions under study}
    	\label{Table1}
    	\begin{tabular}{cclccc}     
    		\hline                   
    		NOAA   & MMC &                  Observation date, & Flux,        & AR         & FI/Max flare \\
                         &   class       &   time (UT)        & 10$^{22}$ Mx & Location   &  (delay)$^1$ \\
    		\hline
    		11734  & A1        &  3 May 2013                 &4.63$\pm$0.42& S18 E12 &3.90/C3.4  \\
     		       &           &(00:00, 06:00, 12:00)   &             &         & (+5.8d)     \\ 
    		12674  & A1        & 4 Sept 2017                 &4.34$\pm$0.15& N13 W00 & 1.41/C5.2   \\
    		       &           &(06:00, 012:00, 18:24)  &             &         &(-4.5d)     \\ 
    		13055  & A1        & 11 Jul 2022                 &4.29$\pm$0.08& S17 E06 & 1.13/C2.9  \\
    		       &           &(06:00, 012:00, 18:24)  &             &         & (+1.0d)     \\ 
    		13282  & A1        & 18 Apr 2023                 &2.74$\pm$0.05& N12 W03 & 3.96/C7.1  \\
    		       &           &(00:00, 06:00, 12:00)   &             &         & (-2.6d)      \\ 
    		\hline
    		11967  & B3        & 3 Feb 2014                  &7.96$\pm$0.15& S13 W00 & 75.03/M6.6  \\
    		       &           &(00:00, 06:00, 12:00)   &             &         & (-3.3d)     \\ 
    		12192  & B3        & 22 Oct 2014                 &13.82$\pm$0.38&S14 W10 & 202.44/X3.1  \\
    		       &           &(00:00, 12:00, 23:48)   &             &         & (+1.9d)     \\ 
    		12371  & B3        & 20 Jun 2015                 &5.91$\pm$0.37& N13 E14 & 21.81/M7.9  \\
    		       &           &(00:00, 06:00, 12:00)   &             &         & (+4.8d)     \\ 
    		12673  & B3        & 5 Sept 2017                 &4.27$\pm$0.04& S08 W17 & 206.87/X9.3  \\
    		       &           &(11:00, 12:00, 13:00)   &             &         & (+1.0d)     \\ 	 
    		\hline
    		\tabnote{Positive(negative) delay in days denotes the time interval between the end (beginning) of the observations and the maximum flare.  }
    	\end{tabular}
    \end{table}
    
 \section{Correlation Functions}
 \label{S-CF} 
 
The correlation function, $Cor(r)$, of a 2D array $u({\bf x})$ is defined as \citep{Monin1971}:
\begin{equation}
 	\label{Eq1}
 	  Cor(r)= \langle(u({\bf x}+{\bf r})-\langle u \rangle)\cdot (u({\bf x})-\langle u \rangle)\rangle, 
 \end{equation}
where ${\bf r}$ is a separation vector, and ${\bf x} \equiv (x,y)$ is the current point within the field of view (FOV). Angle brackets denote averaging over the area. To compare correlation functions of magnetograms with different areas, a normalization of $Cor(r)$ by the variance of the array, $Cor(0)$, was performed. Let us denote the normalized correlation function as $C(r)$:
 \begin{equation}
 	\label{Eq2}   
         C(r) = Cor(r)/Cor(0). 
  \end{equation}
For any 2D data, this function diminished from 1.0 (at $r=0$) as the spatial lag $r$ increases. Here we will focus on $C(r)$ and, for simplicity,  will refer to this function as correlation function, omitting the descriptor ``normalized''. 

Note that a white noise process has a correlation function zero for all $r$ except for $r=0$ when $C(r)=1$, which implies that the process is completely uncorrelated. On the contrary, an array of constant values exhibits the correlation function equal to unity for all lags $r$. In between these two asymptotic cases, the entire variety of physical processes in nature display generally diminishing (or waving) correlation functions allowing us to infer some information about the underlying process. 
 
\begin{figure}
\centerline{\includegraphics[width=1.0\textwidth,clip=]{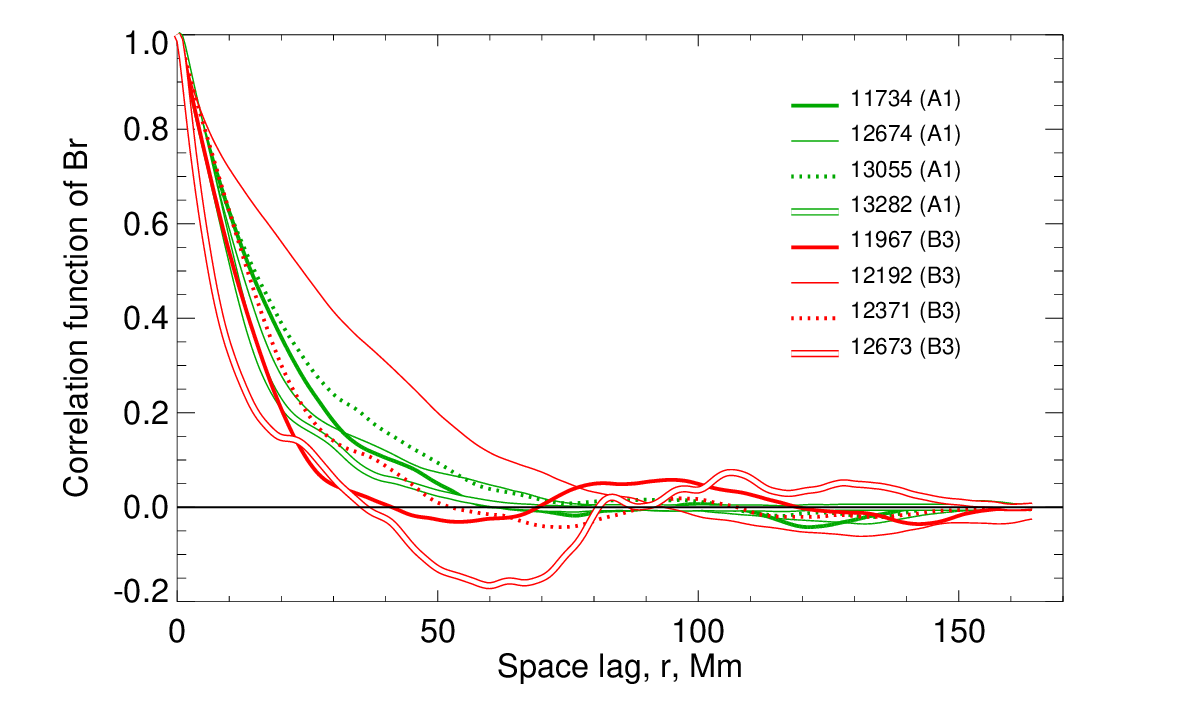}}
\caption{Correlation functions for 8 ARs plotted in the double linear axes.}
\label{fig3}
\end{figure}

For each chosen AR we selected three magnetograms, the exact time of the magnetograms is shown in Table \ref {Table1}. Predominantly, the magnetograms were separated by a six-hour interval, except for the very fast varying AR 12673 and very slow evolving AR 12192.   The correlation function for a given AR was calculated as the average of three correlation functions (Figure \ref{fig3}). While the correlation functions for regular ARs (green curves) are very similar and close to each other, those calculated for irregular ARs (red curves), exhibit broad spread and rather a unique behaviour. Apart from the above qualitative differences, we do not observe any systematic difference in $C(r)$ between A1-class and B3-class ARs as the red and green curves are mixed. It is, therefore, reasonable to apply analytical fits to the $C(r)$, in particular, power law and exponential approximations.  
  
The power law approximation was performed utilizing a function 
 \begin{equation}
 	\label{Eq3}   
 	y=a\cdot x^{\alpha},
 \end{equation}
while the exponential approximation was performed using common logarithm: 
 \begin{equation}
 	\label{Eq4}   
 	y=b\cdot 10^{\beta x}. 
 \end{equation}
Here, $x$ is the spatial scale, $r$, and $y$ is the model function. We used the same scale range, $\Delta r$, for both approximations.  The broadest scale range that the data allowed us to adopt was $\Delta r = 1-36$~Mm, since on large scales $C(r)$ for NOAA AR 12673 becomes negative and the logarithm operation failed.
 
\begin{figure}
\centerline{\includegraphics[width=1.0\textwidth,clip=]{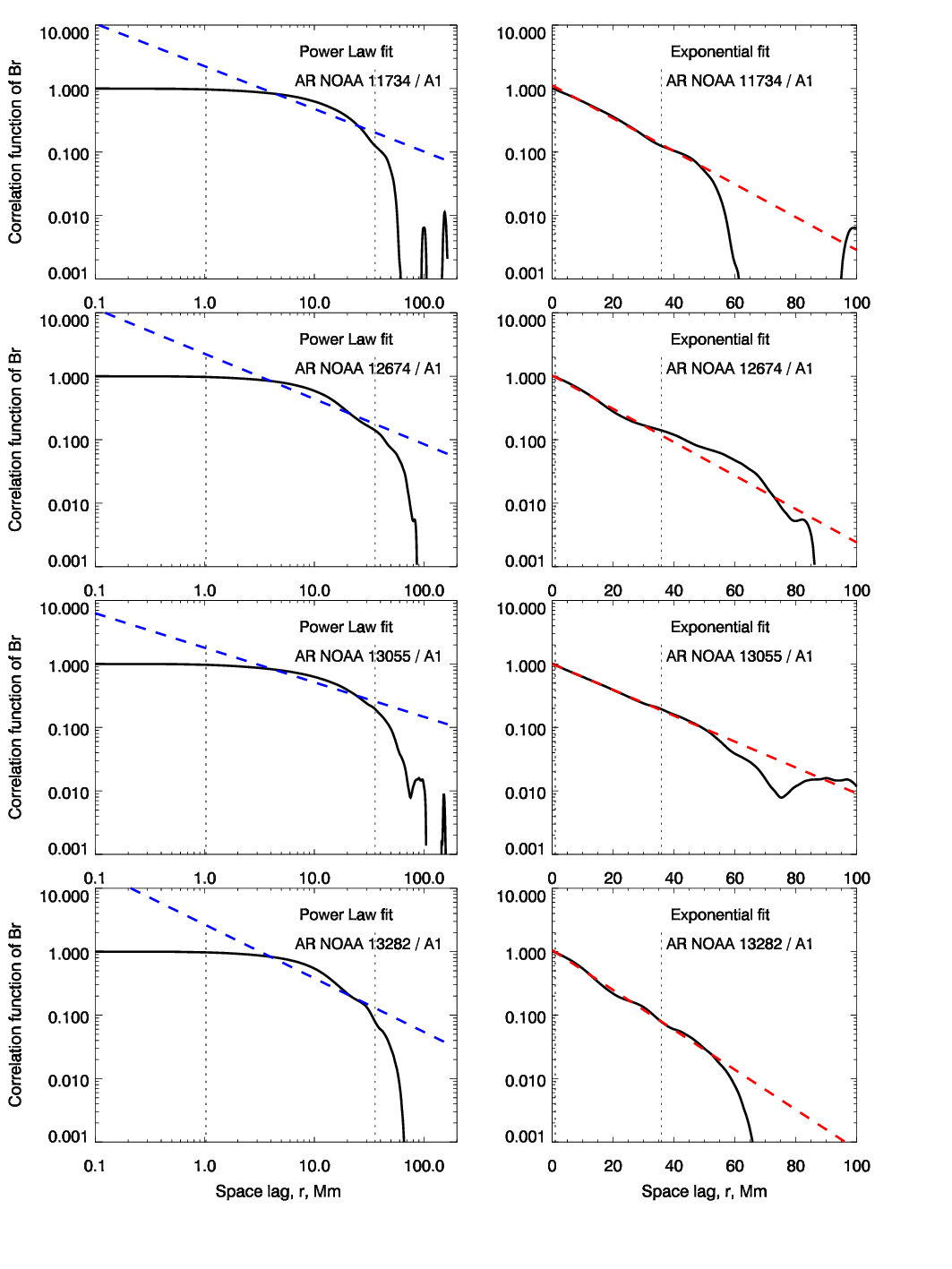}}
\caption{Correlation functions of four regular ARs and their approximations (dashed lines). Left - power law approximations (Equation \ref{Eq3}) plotted in the double-logarithmic coordinates. Right - exponential approximations (Equation \ref{Eq4}) plotted in the linear-logarithmic coordinates. The best linear fit (dashed lines) in each plot was calculated within $1-36$~Mm range marked by the dotted vertical segments.}
\label{fig4}
\end{figure}

\begin{figure}
\centerline{\includegraphics[width=1.0\textwidth,clip=]{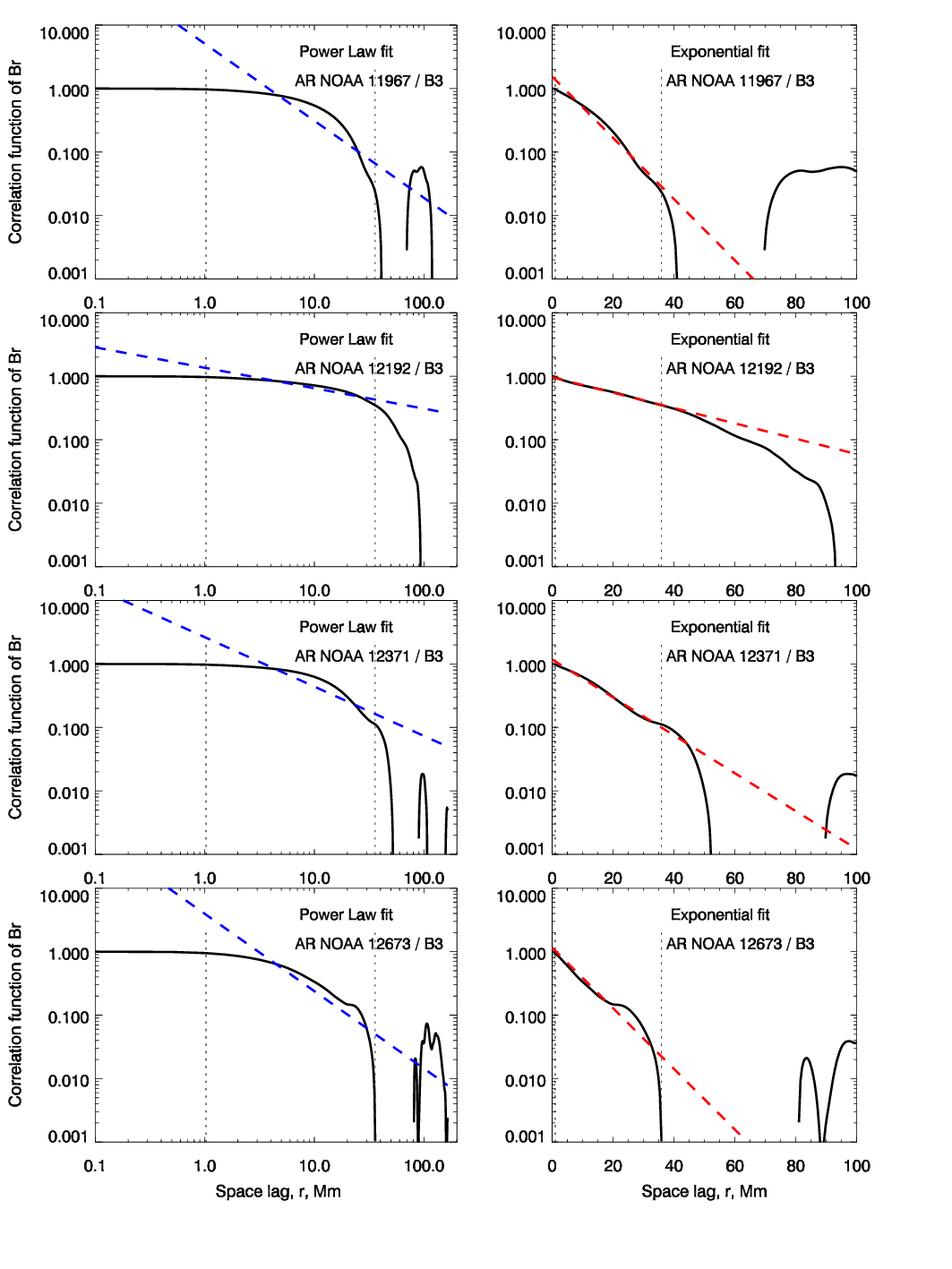}}
\caption{Correlation functions of four irregular ARs and their approximations. Notations are the same is in Figure \ref{fig4}.}
\label{fig5}
\end{figure}

A linear regression was calculated for each AR for both approximations. The results for regular ARs (class A1) are presented in Figure \ref{fig4}, and for irregular ARs (B3-class) in Figure \ref{fig5}. The power law approximation results are shown in the left columns of the figures and the results of the exponential approximations are plotted in the right column of the figures. 
For the power law approximation (Eq \ref{Eq3}), the best linear fit between $lg(r)$ and $lg(C(r))$ was derived. For the exponential approximation (Eq \ref{Eq4}), the same was done between $r$ and $lg(C(r))$. The IDL procedure LINFIT was used, which allows us to derive also the sum of squared errors, $\chi^2$, between the observed $lg(C(r_i))$ and approximated $Y_i \equiv A_0 + A_1 x_i$. For the power law fit, the later expression is:
   \begin{equation}
  	\label{Eq5}   
  	Y_i=lg (a) + \alpha \cdot lg (r_i), 
  \end{equation}  
while the equivalent expression for the exponential fit is:
  \begin{equation}
	\label{Eq6}   
	Y_i=lg (b) + \beta \cdot r_i. 
\end{equation}  
The LINFIT procedure also provides one-sigma errors of the linear fit parameters, which we denote as $\sigma(lg(a)), \sigma  (\alpha), \sigma(lg(b)), \sigma(\beta))$. Parameters of the power law and exponential fits are listed in Tables \ref{Table2} and \ref{Table3}, respectively. 
 
\begin{table}
\caption{Parameters of the power law fit for all ARs}
\label{Table2}
\begin{tabular}{ccccccc}     
\hline                   
 		NOAA    & MMC-class & $lg(a)$  & $\sigma(lg(a))$& $\alpha$ &$\sigma (\alpha)$  & $\chi^2$   \\
 		\hline
 		11734  & A1        & 2.242   & 0.049            & -0.671   &  0.040            &  0.950    \\
 		12674  & A1        & 2.258   & 0.040            & -0.712   &  0.033            &  0.649    \\
 		13055  & A1        & 1.798   & 0.033            & -0.544   &  0.027            &  0.423    \\
 		13282  & A1        & 2.673   & 0.049            & -0.846   &  0.040            &  0.948    \\
 		\hline
 		11967  & B3        & 5.086   & 0.104            & -1.214   &  0.085            &  4.289    \\
 		12192  & B3        & 1.360   & 0.020            & -0.322   &  0.016            &  0.154    \\
 		12371  & B3        & 2.639   & 0.055            & -0.777   &  0.045            &  1.200    \\
 		12673  & B3        & 3.950   & 0.130            & -1.219   &  0.107            &  6.704    \\	
 		\hline
 	\end{tabular}
\end{table}
 
 \begin{table}
	\caption{Parameters of the exponential fit for all ARs}
	\label{Table3}
	\begin{tabular}{ccccccc}     
		\hline                   
		NOAA    & MMC-class & $lg(b)$  & $\sigma(lg(b))$& $\beta$ &$\sigma (\beta)$  & $\chi^2$   \\
		\hline
		11734  & A1        & 0.0501   & 0.00491         & -0.0259   &  0.00023       &  0.0260    \\
		12674  & A1        & 0.0136   & 0.00796         & -0.0263   &  0.00038       &  0.0686    \\
		13055  & A1        & 0.00033  & 0.00172         & -0.0204   &  8e-05         &  0.0032    \\
		13282  & A1        & 0.0254   & 0.00801         & -0.0314   &  0.00038       &  0.0695    \\
		\hline
		11967  & B3        & 0.188   & 0.0182           & -0.0482   &  0.00086       &  0.361    \\
		12192  & B3        & -0.0166 & 0.00158          & -0.0121   &  7e-05         &  0.0027    \\
		12371  & B3        & 0.0702  & 0.00645          & -0.0300   &  0.00031       &  0.0452    \\
		12673  & B3        & 0.0643  & 0.0543           & -0.478    &  0.0026        &  3.194    \\	 	
		\hline
	\end{tabular}
\end{table}

Comparison of the left and right columns in Figures \ref{fig4} and \ref{fig5} shows that the power law failed to approximate the observed correlation functions, whereas the exponential fit appears to be well compatible with data withing the linear range. The power law performed equally poor for both A1- and B3-class ARs, while the exponential fit performed somewhat better for A1-class ARs, where the observed linear range extends further to toward larger $r$ and beyond the vertical dotted line at 36~Mm. The last columns in Tables \ref{Table2} and \ref{Table3} further confirm that the goodness of the exponential fit is better as soon as the $\chi^2$ fitting errors are by the order of magnitude lower than those for the power law. The only exception is the extremely complex NOAA AR 12673, where the exponential law fitting errors are only half of those of the power law fit.      
 
\begin{figure}
\centerline{\includegraphics[width=1.0\textwidth,clip=]{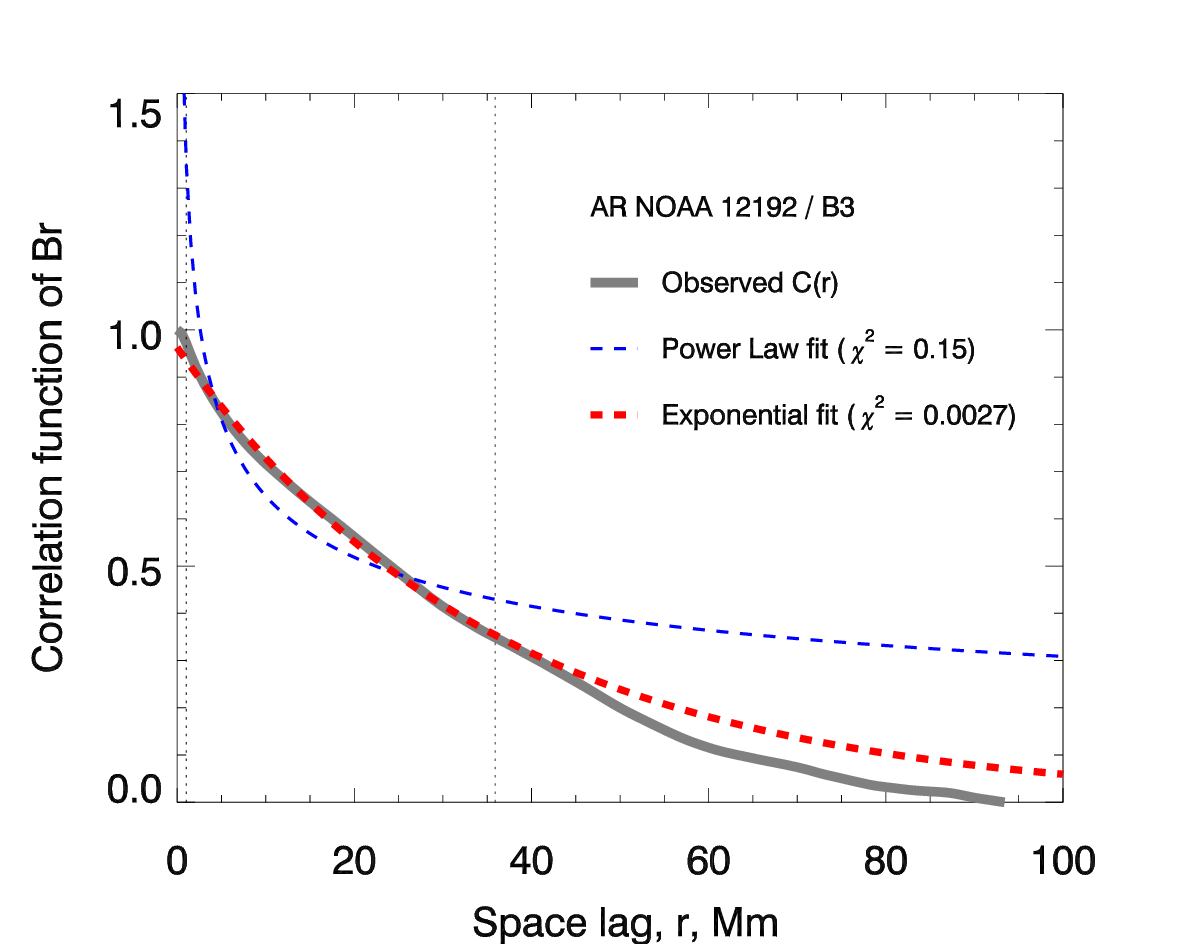}}
\caption{An example of two approximations plotted for NOAA AR 12192 in double-linear coordinates. The vertical dotted line segments mark the region where the approximations were applied.}
\label{fig6}
\end{figure}
 
The superior performance of the exponential fit is further illustrated in Figure \ref{fig6}. The heavy-tailed power law (blue) does not fit the observed correlation function (black). The exponential approximation, while being a good fit at the scales below 50~Mm, only slightly overestimates the observed correlation on larger spatial scales.  Moreover, as it follows from Figure \ref{fig3}, in some cases (e.g., strong $\delta$-structure in NOAA AR 12673)   magnetic correlation functions can show negative values, indicating anti-correlation on large scales. This effect might be caused by the close proximity of opposite polarity magnetic field. 

Based on these results and following \citet{Bak1987} and \citet{Watkins2016}, we may conclude that the photospheric magnetic field is not in the state of SOC, as soon as their correlation function does not obey the power law. On the other hand, \citet{Bak1989} and \citet{Watkins2016} state that "Fractals in nature originate from self-organized {\it critical} dynamical processes". So, it would be interesting to explore the fractal properties of the investigated ARs.  
 
 \section{Fractal Properties of the Magnetic Field in the ARs}
 \label{S-Fr}
 Fractal properties of various objects in nature have been a subject of interest since the famous publication by \citet{Mandelbrot1983}. Generally, a fractal (monofractal), as a self-similar object, may be characterized by one scalar parameter - a fractal dimension (see, e.g., \citet{Feder1988}). In nature we deal with multifractals, which are superpositions of monofractals. In this case, a single scalar parameter is not sufficient to describe such a system, and a spectrum of multifractality was introduced. In solar physics, two ways to explore multifractals were suggested. A multifractality spectrum may be calculated from H$\ddot{o}$lder exponent and Hausdorff dimension as proposed by \citet{McAteer2005, McAteer2007, Conlon2008, McAteer2010}. Another way is based on the structure functions analysis \citep{Monin1971, Frisch1995} that was introduced in \citet{Abramenko2002} and further elaborated in \citet{Abramenko2005a,Abramenko2010}. Here we will use the latter approach.
 
Structure functions were first introduced by \citet{Kolmogorov1941}, and they are defined as statistical moments of field increments (see, e.g., \citet{Monin1971}):
\begin{equation}
\label{Eq7}   
S_q(r)=\langle|u({\bf x}+{\bf r}) - u({\bf x})|^q\rangle,
\end{equation}  
where ${\bf x}$  is a current pixel on a magnetogram,  ${\bf r}$ is a separation vector between any two points used to measure the increment, and $q$ is the order of a statistical moment (a real number).  Angular brackets denote averaging over the area. To derive the spectrum of multifractality, we calculated the ratio of the sixth moment to the cube of the second moment, called a flatness function $F(r)$  \citep{Frisch1995, Abramenko2005a, Abramenko2010}:
\begin{equation}
\label{Eq8}   
F(r)=S_6(r)/(S_2(r))^3  \sim r^{-\kappa}.
\end{equation}   
In the case of a monofractal, the flatness function does not depend on scale (it is flat). On the contrary, for multifractals, $F(r)$  grows (approximately as a power law) when the scale decreases \citep{Frisch1995, Abramenko2005a}. The slope of the flatness function, $\kappa$, determined within the range of the growth, called the flatness function exponent, characterizes the degree of multifractality.

\begin{figure}
\centerline{\includegraphics[width=1.0\textwidth,clip=]{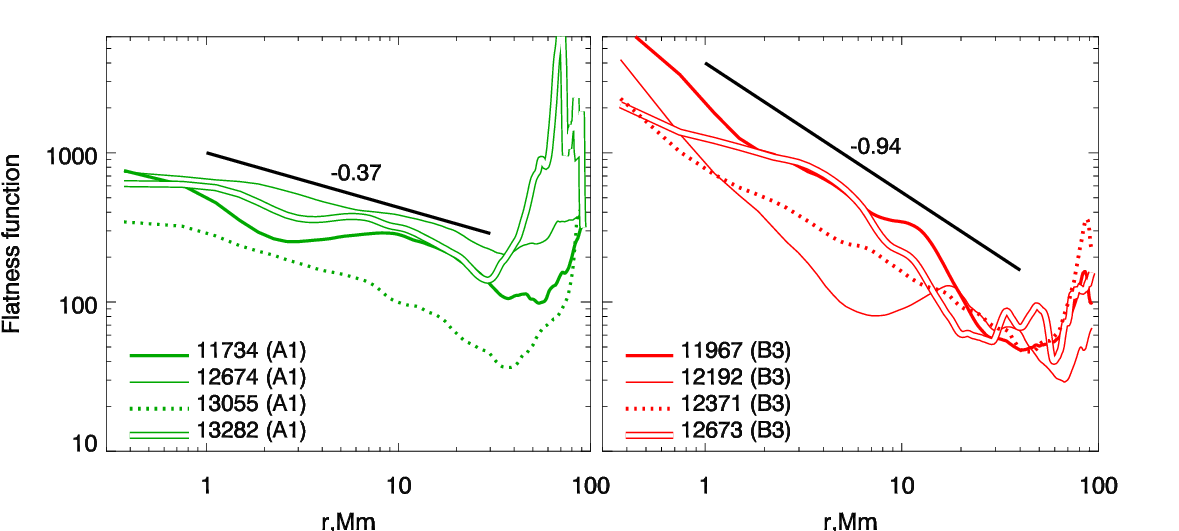}	}
\caption{ Flatness functions calculated by Eqs \ref{Eq7}, \ref{Eq8} for regular (left) and irregular (right) ARs. The power law corresponding to the average slope of the flatness functions of 4 ARs is shown with a black line segment {\bf in each frame}.   }
\label{fig7}
\end{figure}

We calculated flatness functions for all ARs using the algorithm described in \citet{Abramenko2005a, Abramenko2010}. Figure \ref{fig7} shows the result: for A1-class (left panels) and B3-class (right panels) ARs. On scales below approximately 30-50~Mm, a persistent increase of $F(r)$ with the scale decrease is observed for all ARs of both classes. Such a quasi-linear increase in the double-logarithmic plot implies a power law behavior and power law approximations were calculated for each AR over various scale ranges. Table \ref{Table4} shows the results of the power law fitting. As it was shown in \cite{Abramenko2010}, the index $\kappa$ usually varies with the scale range due to individual properties of an AR. Here we estimated the magnitude of $\kappa$ from the broadest range (unique for all ARs, columns 3 and 4 in Table \ref{Table4}), as well as for the narrow range, individual for each AR (columns 5 and 6). For the most complex AR NOAA 12673 we explored two narrow ranges, see the additional bottom line in Table \ref{Table4}.)  The average index $\kappa$ was found to be -0.37$\pm$0.10  from eight estimates of the A1-class ARs, and much higher index of $\kappa=-0.94\pm$0.36 was found for (nine estimates of) the B3-class ARs. The indices and the power law segments are shown in Figure \ref{fig7}.

 \begin{table}
	\caption{ {\bf Parameters of the fitting of the flatness functions for all ARs}}
	\label{Table4}
	\begin{tabular}{cccccc}     
		\hline                   
		NOAA    & MMC-class & Broad range, Mm  & $\kappa$& Narrow range, Mm &$\kappa$     \\
		\hline
		11734  & A1        & 0.4 - 29   & -0.289        & 5.6 - 19  &  -0.276           \\
		12674  & A1        & 0.4 - 29   & -0.301        & 1.1 - 19  &  -0.272           \\
		13055  & A1        & 0.4 - 29   & -0.542        & 0.7 - 5.6 &  -0.386           \\
		13282  & A1        & 0.4 - 29   & -0.382        & 5.6 - 19  &  -0.478           \\
		\hline
		11967  & B3        & 0.4 - 29  & -1.083         & 0.4 - 17  &  -0.842           \\
		12192  & B3        & 0.4 - 29  & -0.557         & 0.4 - 6.3 &  -1.381           \\
		12371  & B3        & 0.4 - 29  & -0.737         & 1.9 - 29  &  -0.732           \\
		12673  & B3        & 0.4 - 29  & -1.097         & 0.4 - 5.2 &  -0.449           \\	
		12673  &           &           &                & 3.7 - 22  &  -1.554           \\ 	
		\hline
	\end{tabular}
\end{table}

We thus confirmed that both types of ARs, simple bipolar A1-class ARs and vary complex multipolar B3-class ARs, display the property of multifractality, which is more pronounced for complex ARs. In general, this conclusion is in a good agreement with previous studies by \citet{Abramenko2005a,McAteer2007, McAteer2010,Abramenko2010}. The aim of this test was to confirm that the studied ARs, being observed with another instrument, also exhibit properties of multifractality.

 \section{Concluding Remarks}
 \label{S-Con}

We explored properties of the radial component of photospheric magnetic fields, $B_r$, in ARs. The analysis was performed for two very different types of ARs: i) simple bipolar magnetic structures with regular orientation (MMC-class A1) and ii) very complex multipolar ARs (MMC-class B3). We calculated spatial correlation functions and their analytical approximation which showed that the exponential approximation, determined in the scale range of 1-36~Mm is the best fit for the correlation functions of both types ARs, while the power law has failed to approximate the observed correlation functions. Additional investigation of fractal properties of the same data revealed all studied ARs exhibit multifractal property, although they are more pronounced in case of the complex B3-class ARs. 

Correlation functions are not a widely used tool for investigating spatial structures in solar physics mainly because it requires high-resolution and high accuracy measurements of photospheric magnetic fields acquired over large areas, as well significant computational efforts. Meanwhile, according to \citet{Aschwanden2016, Watkins2016, McAteer2016}, correlation functions can be used as a powerful diagnostic tool that may reveal dependence of data on various spatial separation lags. Their basic role in interpreting short- and long-distance connections is discussed in details in \citet{McAteer2016}. In particular, white noise is a short-correlation array and it has a non-zero correlation only at the zero lag. On the contrary, for a system capable of spontaneously producing extremely large fluctuations, the correlations must be large for large lags. This property is usually referred to as SOC. That is why \citet{Watkins2016}, based on the classical SOC concept \citep{Bak1989}, argued that correlation functions of SOC-systems must be power law functions rather than exponential ones. Following \citet{Bak1989}, \citet{Watkins2016} further argued that fractals are the result of the dynamical development of a system in the state of SOC, in other words, fractality is a necessary consequence of the SOC state. However, \citet{Aschwanden2018} defined SOC-systems as a subset of systems with self-organization, and fractality is assumed to be the property of all systems with self-organization and the state of SOC is not a necessary condition for fractality.
     
Based on the above reasoning, our results allow us to conclude that photospheric magnetic fields in solar ARs represent a self-organized system, which, however, is not in the state of SOC. This conclusion does not depend on the complexity of ARs. At the same time, these magnetic fields display multifractal properties, which leads us to conclude that fractality is an attribute of self-organized systems as well and not only of systems with SOC.   

Although the above considerations refer to general issues, they also closely relate to the problems of AR origin. To date, it is undoubtedly established that the magnetic field is an energy source and structural skeleton for non-stationary processes in the solar atmosphere occurring on a wide range of spatial and temporal scales spanning from nano-flares to extremely large coronal mass ejections. The SOC state has been confirmed for temporal processes in the corona (see, e.g., \citet{Aschwanden2016, McAteer2016} and references herein), allowing us to suggest that the magnetic field in the corona is in the SOC state. At the same time, the magnetic field is only self-organized in the photosphere. A question arises then, how the critical state in the corona is achieved and how is it connected to the observed self-organization in the photosphere? Ultimately, this question leads to a problem of coupling between the photosphere and the corona.

\section{Acknowledgments}
We are thankful to anonymous referee whose comments helped much to improve the paper.
SDO is a mission for NASA {\it Living With a Star} (LWS) program. The SDO/HMI data were provided by the {\it Joint Science Operation Center} (JSOC).

\begin{authorcontribution}
VIA wrote the manuscript, RAS prepared data, figures and tables, both authors contributed to the analysis and reviewed the manuscript.
\end{authorcontribution}

\begin{fundinginformation}
	ddd
\end{fundinginformation}
No funding
\begin{dataavailability}
The SDO/HMI data are available via the {\it Joint Science Operation Center} (JSOC). The data obtained in the paper can be offered by the authors by request.  
\end{dataavailability}

\begin{ethics}
	\begin{conflict}
The authors declare no conflict of interests.
\end{conflict}
\end{ethics}

 {}

\end{article} 


\begin{thebibliography}{}
 	
 \bibitem[Abramenko \emph{et al.}(2002)]{Abramenko2002}Abramenko, V.I., Yurchyshyn, V.B., Wang, H., Spirock, T.J., and Goode, P.R.: 2002, {\it The Astrophysical Journal} {\bf 577}, 487. doi:10.1086/342169.
 
 \bibitem[Abramenko \emph{et al.}(2003)]{Abramenko2003}Abramenko, V.I., Yurchyshyn, V.B., Wang, H., Spirock, T.J., and Goode, P.R.: 2003, {\it The Astrophysical Journal} {\bf 597}, 1135. doi:10.1086/378492.
 
 \bibitem[Abramenko(2005a)]{Abramenko2005a}Abramenko, V.I.: 2005a, {\it Solar Physics} {\bf 228}, 29. doi:10.1007/s11207-005-3525-9.
 
 \bibitem[Abramenko(2005b)]{Abramenko2005b}Abramenko, V.I.: 2005b, {\it The Astrophysical Journal} {\bf 629}, 1141. doi:10.1086/431732.
 
 \bibitem[Abramenko and Yurchyshyn(2010)]{Abramenko2010}Abramenko, V. and Yurchyshyn, V.: 2010, {\it The Astrophysical Journal} {\bf 722}, 122. doi:10.1088/0004-637X/722/1/122.
 	
 \bibitem[Abramenko, Zhukova, and Kutsenko(2018)]{Abramenko2018}Abramenko, V.I., Zhukova, A.V., and Kutsenko, A.S.: 2018, {\it Geomagnetism and Aeronomy} {\bf 58}, 1159. doi:10.1134/S0016793218080224.
 	
 \bibitem[Abramenko(2021)]{Abramenko2021}Abramenko, V.I.: 2021, {\it Monthly Notices of the Royal Astronomical Society} {\bf 507}, 3698. doi:10.1093/mnras/stab2404.
 
 \bibitem[Abramenko, Suleymanova, and Zhukova(2023)]{Abramenko2023}Abramenko, V.I., Suleymanova, R.A., and Zhukova, A.V.: 2023, {\it Monthly Notices of the Royal Astronomical Society} {\bf 518}, 4746. doi:10.1093/mnras/stac3338. 
 
 \bibitem[Aschwanden \emph{et al.}(2016)]{Aschwanden2016}Aschwanden, M.J., Crosby, N.B., Dimitropoulou, M., Georgoulis, M.K., Hergarten, S., McAteer, J., and, ...: 2016, {\it Space Science Reviews} {\bf 198}, 47. doi:10.1007/s11214-014-0054-6.
 	
 \bibitem[Aschwanden \emph{et al.}(2018)]{Aschwanden2018}Aschwanden, M.J., Scholkmann, F., B{\'e}thune, W., Schmutz, W., Abramenko, V., Cheung, M.C.M., and, ...: 2018, {\it Space Science Reviews} {\bf 214}, 55. doi:10.1007/s11214-018-0489-2.
 	
 \bibitem[Bak, Tang, and Wiesenfeld(1987)]{Bak1987}Bak, P., Tang, C., and Wiesenfeld, K.: 1987, {\it Physical Review Letters} {\bf 59}, 381. doi:10.1103/PhysRevLett.59.381.
 	
 \bibitem[Bak and Chen(1989)]{Bak1989}Bak, P. and Chen, K.: 1989, {\it Physica D Nonlinear Phenomena} {\bf 38}, 5. doi:10.1016/0167-2789(89)90166-8.
 
 \bibitem[Baumgartner \emph{et al.}(2022)]{Baumgartner2022}Baumgartner, C., Birch, A.C., Schunker, H., Cameron, R.H., and Gizon, L.: 2022, {\it Astronomy and Astrophysics} {\bf 664}, A183. doi:10.1051/0004-6361/202243357.
 
 	
 \bibitem[Biskamp(1993)]{Biskamp1993}Biskamp D. Nonlinear Magnetohydrodynamics. Cambridge: Cambridge University Press; 1993. pp 378. doi:10.1017/CBO9780511599965
 	
 \bibitem[Conlon \emph{et al.}(2008)]{Conlon2008}Conlon, P.A., Gallagher, P.T., McAteer, R.T.J., Ireland, J., Young, C.A., Kestener, P., and, ...: 2008, {\it Solar Physics} {\bf 248}, 297. doi:10.1007/s11207-007-9074-7.
 	
 \bibitem[Frisch(1995)]{Frisch1995}Frisch U. Turbulence: The legacy of A.N. Kolmogorov. Cambridge: Cambridge University Press; 1995. pp 296. ISBN 0-521-45713-0
 	
 \bibitem[Feder(1988)]{Feder1988} Feder, J.:1988, The Fractal Dimension. In: Fractals. Physics of Solids and Liquids. Springer, Boston, MA. https://doi.org/10.1007/978-1-4899-2124-62
 	
 	
 \bibitem[Kolmogorov(1941)]{Kolmogorov1941}Kolmogorov, A.: 1941, {\it Akademiia Nauk SSSR Doklady} {\bf 30}, 301.
 	
 \bibitem[Kurakin (2011)]{Kurakin2011}Kurakin A. : 2011,{\it Theoretical Biology and Medical Modelling} {\bf 8}, 4. http://www.tbiomed.com/content/8/1/4

\bibitem[Mandelbrot(1983)]{Mandelbrot1983}Mandelbrot, B.B.: 1983, {\it New York, W.H. Freeman and Co., 1983, 495 p.}.

\bibitem[McAteer, Gallagher, and Ireland(2005)]{McAteer2005}McAteer, R.T.J., Gallagher, P.T., and Ireland, J.: 2005, {\it The Astrophysical Journal} {\bf 631}, 628. doi:10.1086/432412.

\bibitem[McAteer \emph{et al.}(2007)]{McAteer2007}McAteer, R.T.J., Young, C.A., Ireland, J., and Gallagher, P.T.: 2007, {\it The Astrophysical Journal} {\bf 662}, 691. doi:10.1086/518086.

\bibitem[McAteer, Gallagher, and Conlon(2010)]{McAteer2010}McAteer, R.T.J., Gallagher, P.T., and Conlon, P.A.: 2010, {\it Advances in Space Research} {\bf 45}, 1067. doi:10.1016/j.asr.2009.08.026.

\bibitem[McAteer \emph{et al.}(2016)]{McAteer2016}McAteer, R.T.J., Aschwanden, M.J., Dimitropoulou, M., Georgoulis, M.K., Pruessner, G., Morales, L., and, ...: 2016, {\it Space Science Reviews} {\bf 198}, 217. doi:10.1007/s11214-015-0158-7.

\bibitem[Monin and I'aglom(1971)]{Monin1971}Monin, A.S. and I'aglom, A.M.: 1971, {\it Statistical fluid mechanics; mechanics of turbulence, by Monin, A. S.; I'aglom, A. M. Cambridge, Mass., MIT Press [c1971-75]}.

\bibitem[Scherrer \emph{et al.}(2012)]{Scherrer2012}Scherrer, P.H., Schou, J., Bush, R.I., Kosovichev, A.G., Bogart, R.S., Hoeksema, J.T., and, ...: 2012, {\it Solar Physics} {\bf 275}, 207. doi:10.1007/s11207-011-9834-2.

\bibitem[Schou \emph{et al.}(2012)]{Schou2012}Schou, J., Scherrer, P.H., Bush, R.I., Wachter, R., Couvidat, S., Rabello-Soares, M.C., and, ...: 2012, {\it Solar Physics} {\bf 275}, 229. doi:10.1007/s11207-011-9842-2.

\bibitem[Suleymanova \emph{et al.}(2023)]{Suleymanova2023} Suleymanova, R.A., Miroshnichenko, L.I., Abramenko, V.I.: 2023, {\it Solar Physics} {\bf in press}.                 

\bibitem[Suleymanova, Miroshnichenko, and Abramenko(2024)]{Suleymanova2024}Suleymanova, R.A., Miroshnichenko, L.I., and Abramenko, V.I.: 2024, {\it Solar Physics} {\bf 299}, 7. doi:10.1007/s11207-023-02248-w.

\bibitem[Watkins \emph{et al.}(2016)]{Watkins2016}Watkins, N.W., Pruessner, G., Chapman, S.C., Crosby, N.B., and Jensen, H.J.: 2016, {\it Space Science Reviews} {\bf 198}, 3. doi:10.1007/s11214-015-0155-x.


\end{thebibliography}
\end{document}